# The Death of the Legal Author
# Authority, Intention, And Law-Creation in the Advent of GenAI

Julieta A. Rabanos[*], Bojan Spaić[**]

**Abstract.** Generative artificial intelligence in the form of chatbots based on large language models (LLMs) has taken the world of law by storm. Philosophy of law is struggling to catch up with the theoretical significance of the advent of technological development and the way it may modify traditionally established understanding of legal phenomena, such as law-creation and authority. In this sense, for the most part, heated philosophical debates have circled around a normative question: "*Should* AI create and interpret law?". Much less attention has been given to a different, albeit previous, question: "C*an* AI create and interpret law?" That is, is AI capable of producing outputs that can be deemed as "law" (at least, law as we know it)? Can AI be a "legal author"?

This paper explores this unattended question and endeavours to provide some provisional answers. In the first part, we define legal authority, legal authorship, and legal interpretation and claim that the intention of a determinate authoritative author is often considered the condition of the possibility of creating and interpreting contemporary legal texts. In the second part, we argue that LLM AI, in general, and ChatGPT, in particular, generate legal texts without having any intention. In the third part, we consider the positions of the authors that downplay or even eliminate intention from the discussions about the legal character of prescriptive texts. 4. Finally, we argue that there are good reasons to side with the second group of authors. The ability of agents without intentions, like ChatGPT, to create legal text is an argument in favour of the thesis that law can be created without intention behind the creation and that nonintentional creation can be interpreted to arrive at legal norms.

[*] Faculty of Law University of Belgrade, julieta.rabanos@ius.bg.ac.rs
[**] Faculty of Law University of Belgrade, bojan.spaic@ius.bg.ac.rs

## I. INTRODUCTION

In the last week of November 2023, Ramiro Rosário, a city councilman in Porto Alegre (Brazil), revealed that on 23 November, an entire legislative act adopted by the city council consisting of 35 councilmen had been written by ChatGPT. Rosário reported entering a 49-word prompt into ChatGPT, which returned the full draft proposal within seconds, including justifications.[1] At the beginning of the same year, Barry Finegold, a state senator in Massachusetts (United States), turned to ChatGPT to help write a bill to regulate artificial intelligence models, including ChatGPT. The bill included a Section 3 containing the following disclaimer: "*This act has been drafted with the help of ChatGPT and any errors or inaccuracies in the bill should not be attributed to the language model but rather to its human authors*".[2] In July 2023, it was part of a hearing of the Joint Committee on Advanced Information Technology, the Internet and Cybersecurity[3], and at the end of 2023, it was incorporated as part of a new bill (an "emergency law") reported by that Committee. Up to this day, this bill is yet to be voted on.[4]

These recent cases are representative of what starts to be a worldwide trend regarding generative artificial intelligence and law. In the last decade, AI has been consistently and increasingly used in legal practice for various purposes. Some of its current uses include 1) document review (AI algorithms can be trained to automatically review and classify large volumes of legal documents such as contracts, leases, and legal briefs)[5]; 2) smart contracts (computer programs that automatically execute the terms of a contract when certain conditions are met)[6]; 3) legal research (AI algorithms can be used to search and analyse legal precedents, case law, and legislation)[7]; 4) due diligence (AI can help to identify potential legal risks and compliance issues during mergers and acquisitions)[8]; 5) predictive analytics (AI can analyse large volumes of data to identify patterns and make predictions about legal outcomes)[9]; and 6) chatbots (AI-powered chatbots can be used to provide basic legal information and answer common questions)[10]. Recent developments in artificial intelligence have made it possible to use AI in law not only as a tool – an aid in the daily tasks of lawyers – but, with certain caveats, also as a semi-autonomous drafter of briefs, contracts, and even laws, a partner legal researcher and even a legal problem solver.

---

[1] See: https://www.euronews.com/next/2023/12/03/politicians-in-a-city-in-brazil-passed-a-law-secretly-written-by-chatgpt-and-now-there-is-z|https://apnews.com/article/brazil-artificial-intelligence-porto-alegre-5afd1240afe7b6ac202bb0bbc45e08d4 | Reportedly, he hid the origin of the document because had he "revealed it before, the proposal certainly wouldn't even have been taken to a vote," and that it "would be unfair to the population to run the risk of the project not being approved simply because it was written by artificial intelligence". See: https://www.seattletimes.com/business/brazilian-city-enacts-an-ordinance-secretly-written-by-a-surprising-new-staffer-chatgpt/

[2] See: https://malegislature.gov/Bills/193/S31/Senate/Bill/Text | Reportedly, Finegold used ChatGPT in the drafting to "illustrate its power", as in his opinion the use of AI help draft new legislation is inevitable, but also wanted the origins of the content to be explicit: "We want work that is ChatGPT generated to be watermarked, (...) I'm in favor of people using ChatGPT to write bills as long as it's clear." See: https://www.seattletimes.com/business/brazilian-city-enacts-an-ordinance-secretly-written-by-a-surprising-new-staffer-chatgpt/

[3] See: https://malegislature.gov/Events/Hearings/Detail/4626

[4] See: https://malegislature.gov/Bills/193/S2539

[5] See e.g. Dietrich Trautmann, "Large Language Model Prompt Chaining for Long Legal Document Classification" (2023). Available at https://doi.org/10.48550/arXiv.2308.04138

[6] See e.g. Mateja Djurovic & André Janssen, "The Formation of Blockchain-based Smart Contracts in the Light of Contract Law", *European Review of Private Law* 26 (2018): 753-772; Mateja Djurovic & André Janssen, Formation of Smart Contracts under Contract Law, in Larry DiMatteo, Michel Cannarsa, & Cristina Poncibò (eds), *The Cambridge Handbook of Smart Contracts, Blockchain Technology and Digital Platforms* (Cambridge: Cambridge University Press, 2019): 61-79.

[7] See e.g. Jonathan Choi, "How to Use Large Language Models for Empirical Legal Research", *Journal of Institutional and Theoretical Economics (JITE)* 180 (2024): 214-233; Serena Villata, Michal Araszkiewicz, Kevin Ashley *et al,* "Thirty years of *artificial intelligence and law: the third decade*", *Artificial Intelligence and Law* 30 (2022): 561–591.

Admittedly, the full potential of current-generation AI to change the practices of the three basic law-related operations (law-creation, law-application, and law-execution) within contemporary democracies is yet to be determined. Still, the willingness of legal practitioners to use the commercially available models to aid them in those tasks is even now very well documented. Other than the cited cases of law-creation in Brazil and the United States, ChatGPT has been used to argue in front of a court of law.[11] Civil servants and even governments[12] use it in communication with citizens and automation of repetitive tasks[13], with rules already being issued that limit the use in the public sector[14].

In the face of all the technological developments, made possible by the advent of transformer models of AI, several positions have been adopted across the board in legal theory and philosophy. On the one hand, there are those who strongly support the use of AI and LLMs for traditional legal tasks, even arriving to support the gradual replacement of human intervention in them (at least, the most menial ones)[15]. Reasons for this support vary from efficiency and elimination of human biases to superior epistemological capacities by means of superior computational capacities. On the other hand, there are those who raise strong concerns not only about the current and eventual usage of AI and LLMs for traditional tasks but especially about the possibility of replacing humans in most legal tasks.[16] Reasons for concerns vary from current limitations in technology resulting in "hallucinations" and reproduction of human biases under the guise of objectivity to the impossibility of ascribing responsibility to an entity that does not make "decisions" in the traditional sense.

[8] See e.g. David Krause, "ChatGPT and Other AI Models as a Due Diligence Tool: Benefits and Limitations for Private Firm Investment Analysis" (April 11, 2023). Available at https://ssrn.com/abstract=4416159

[9] See e.g. Karolina Mania, "The Digital Transformation of Legal Industry: Management Challenges and Technological Opportunities", *DANUBE* 13 (2022): 209-225.

[10] See e.g. Marc Queudot, Éric Charton, Marie-Jean Meurs, "Improving Access to Justice with Legal Chatbots", *Stats* 3 (2020):356-375; Andrew Perlman, "The implications of ChatGPT for legal services and society" (December 11, 2022), *Suffolk University Legal Studies Research Paper Series* n. 22-14. Available at https://ssrn.com/abstract=4294197

[11] See: https://www.forbes.com/sites/mollybohannon/2023/06/08/lawyer-used-chatgpt-in-court-and-cited-fake-cases-a-judge-is-considering-sanctions/?sh=48feca477c7f

[12] See: https://govinsider.asia/intl-en/article/chatgpt-and-the-public-service

[13] See: https://www.government-transformation.com/innovation/japan-to-use-chatgpt-as-a-tool-in-city-government

[14] See: https://www.cbc.ca/news/politics/generative-ai-chatgpt-government-1.6961323

[15] See e.g. Jaromir Savelka *et al*, "Can GPT-4 Support Analysis of Textual Data in Tasks Requiring Highly Specialized Domain Expertise", *ITiCSE 2023: Proceedings of the 2023 Conference on Innovation and Technology in Computer Science Education V. 1* (2023): 117-123, available at https://doi.org/10.48550/arXiv.2306.13906; Guilherme da Franca Couto Fernandes de Almeida *et al*, "Exploring the psychology of LLM's Moral and Legal Reasoning", *Artificial Intelligence* 333 (2024): 104145; Lauren Martin *et al*, "Better Call GPT, Comparing Large Language Models Against Lawyers" *arXiv* (2024). Available at https://doi.org/10.48550/arXiv.2401.16212; Zhiwei Fei *et al*, "LawBench: Benchmarking Legal Knowledge of Large Language Models", *arXiv* (2023). Available at https://doi.org/10.48550/arXiv.2309.16289; Jiaxi Cui *et al*, "ChatLaw: Open-Source Legal Large Language Model with Integrated External Knowledge Bases", *arXiv* (2023). Available at https://doi.org/10.48550/arXiv.2306.16092

For the most part, these positions and the related heated debates circle around a normative question: "*Should* AI create and interpret law?" Much less attention has been given to a different, albeit previous, question: "C*an* AI create and interpret law?" In other words: is AI capable of producing outputs that can be deemed as "law" (at least, law as we know it)? Can AI be a "legal author"?

One way to answer these questions is *descriptive*, entailing extensive testing of commercially available large language models. To a degree, this testing was already conducted on human[17] and machine[18] benchmarks. In a systematic and rigorous way, the research confirms the general feeling of the practical usefulness of ChatGPT and similar commercially available products for lawyers – large language models can, in fact, produce outputs that are convincingly legal and display all signs of what we usually call legal reasoning. The other way of answering these questions is *conceptual*. Namely, philosophical notions of creation and interpretation of law are based on certain fundamental premises. These premises contain the conditions of possibility for something to count as 'law' and 'law-creation'. For one, according to most, contemporary law-creation entails that there is a legal author in a position of authority to create text that is legal in nature. Secondly, we tend to understand the legal product of the determinate legal author as an artefact – a text or sign in which the intention of the authority was materialised and brought to life, with the purpose of regulating behaviour. To even start discussing whether AI can create law and whether AI-made law can be interpreted, it is necessary to establish whether AI fulfils the conditions for the possibility of law-creation, at least in the way these are dominantly conceived in legal theory and philosophy.

This paper will explore this unattended question and endeavour to provide some provisional conceptual answers. In Section II, we will analyse that does it mean to be a creator of law or legal text (what does it mean to be a legal author-ity)[19] and what does it mean for something to count as a law or legal text. A special focus will be on evaluating whether there is a possible understanding of "law" and "law-creation" that does not involve intentionality and/or that does not give foremost importance to the figure of the "creator". In Section III, we will explore large language model AI, their functioning and their "legal" outputs. In Section IV, we will analyse whether the output produced by LLMs could be considered as "law" – and its production, thus, "law-creation" – and under which conditions (if any). In Section V, we will offer some concluding remarks.

[16] See e.g. A. Feder Cooper *et al*, "Report of the 1st Workshop on Generative AI and Law", *arXiv* (2023). Available at https://doi.org/10.48550/arXiv.2311.06477; Samuel Dahan *et al*, "Lawyers Should Not Trust AI", *Queen's University Legal Research Paper* (2023). Available at https://ssrn.com/abstract=4587092; Perlman, "The implications of ChatGPT for legal services and society"; Christoph K. Winter, "The Challenges of Artificial Judicial Decision-Making for Liberal Democracy", in Piotr Bystranowski, Bartosz Janik, Maciej Próchnicki (eds), *Judicial Decision-Making* (Cham: Springer International Publishing, 2022): 179-204; Peter Bernard Ladkin, "Involving LLMs in legal processes is risky", *Digital Evidence and Electronic Signature Law Review* 20 (2023): 40-46; Robert Diab, "Too Dangerous to Deploy? The Challenge Language Models Pose to Regulating AI in Canada and the EU", *University of British Columbia Law Review (forthcoming)* (2024). Available at https://ssrn.com/abstract=4680927; Matthew Dahl *et al*, "Large Legal Fictions: Profiling Legal Hallucinations in Large Language Models", *Journal of Legal Analysis* 16 (2024): 64-93.

[17] See Daniel Martin Katz *et al*, "GPT-4 Passes the Bar Exam", *Philosophical Transactions of the Royal Society A* 382 (2024): 1-17.

[18] See Neel Guha *et al*, "LegalBench: A collaboratively built benchmark for measuring legal reasoning in large language models", *Conference on Neural Information Processing Systems, Datasets and Benchmarks Track, Osgoode Legal Studies Research Paper No. 4583531*. Available at https://ssrn.com/abstract=4583531; Zhiwei Fei *et al*, "LawBench: Benchmarking Legal Knowledge of Large Language Models".

## II. LEGAL AUTHOR(ITY): FIRST TAKE

### *II.1. The Authority: A brief overview of authority and law.*

When discussing authority and law, there are several interconnected threads of questions to tackle and several diverse faces of the same issue to begin with. To mention a handful of them that could come in handy here: the different questions and answers regarding the authority *of* law (commonly connected with the slippery notions of "normativity" and "obedience") and authority *in* law (commonly associated with issues of institutional design and power-conferring rules); the question whether the central notion for the analysis is "authority" or "authoritative", *i.e.* whether a subject is an authority and then their created outputs are thus authoritative, or whether an output is authoritative and then its creator is thus an authority; and different levels of analysis, encompassing normative, descriptive, and conceptual ways to approach the issue and come up with answers. For the purposes of tackling our main question, one fruitful way to do so is to consider the issue of authority and law from the standpoint of law-creation. Hence, here we can interest ourselves in authority as the issue of authorship of certain objects that have certain characteristics and that have been produced by certain means.[20] Let us look at this more closely.

It is quite commonplace in contemporary philosophy of law, following Joseph Raz, to conceive authority in terms of the so-called "claim to legitimate authority"[21]: that is, in terms of a subject issuing or uttering certain prescriptive statements (directives) with the claim of being authorised to do so and of being justified in demanding them to be taken as protected reasons for action by their addressees[22]. Using this as a departing point for reconstruction, it seems that there is a subject[23] (let's call it here "author") with the capacity and/or power to issue or utter statements that can (in some way) be deemed as prescriptive or imperative, and that has (or can be taken as having) issued or uttered those statements with the intention of them being taken as a very specific type of reason for action by other subjects.[24]

---

[19] It is commonplace to consider that law can be created both by a specific, timely-identifiable, intentional act and by a myriad of non-specific, non-timely-identifiable, unintentional acts. This is the traditional distinction between legislation and customs. For present purposes, here we will mainly explore the first way of law-creation (thus focusing on the figure of the "creator"), even if we will come back to the second way in later sections of the paper when discussing the possibility of law-creation (in the first way) without intention.

[20] This has been, in fact, the classical approach to authority; see, for a discussion, e.g. Karl Olivecrona, *Law as fact,* 2 ed (London: Stevens, 1971). Curiously, it also corresponds to the alleged origin of the etymology of the word in English: from the Latin *auctor*, "author" to *autoritas* and *auctoritatem,* "invention, influence, command", and then from Old French *au(c)torité*, "authority, right, permission" to Middle English "*authority*". See e.g. https://www.oed.com/dictionary/authority_n?tl=true&tab=etymology

[21] It is plausible to read Raz as aiming to offer a discourse on the concept of authority *tout court*, where he considers that all types of authority that can be known are in fact instances of the same concept and therefore share the same basic structure. This means that very different kinds such as parental authority or governmental authority, epistemic authority or practical authority, can be understood using the same structure (and having the same basic structural principle of legitimacy related to their function to subjects), and differ only in the identity of already present or additional elements. In this view, legal authority can me conceived a type of political authority, which would be in turn a type of practical authority. See Joseph Raz, *The Authority of Law* (Oxford: Clarendon Press, 1979) and Joseph Raz, *The Morality of Freedom Law* (Oxford: Clarendon Press, 1986).

The elements of this first definition that are relevant for present purposes are four: (1) the "relational" quality of authority; (2) the "capacity" of the subject; (3) the "type of utterances"; and (4) the "intention" or volition behind the act of issuing or uttering. When specifically considering authority *in* law, an additional element entails (5) the "norm-dependence" status of "authority" within a certain legal system.

(1) Regarding the "relational" quality of authority, it is twofold: first, authority exists in relation to subjects (relation to whom); second, it operates within a scope (relation to what). Leaving the latter aside for now, the former means that every authority exists within an authority relation: one composed by the authority and the addressee(s), mediated (if any) by the presence of statements uttered by one of the subjects. The exact articulation of the relation between the parts, their characteristics and their roles in it is *per se* a matter of open discussion. We will come back to this soon.

(2) Regarding the "capacity", albeit that of "claiming legitimate authority" within his specific conception of legitimate authority, Joseph Raz distinguishes between normative and material conditions. For present purposes, the latter seem to be of interest here.[25] These material conditions are related to the possibility of the claiming subject to actually perform its service or its role as an authority: at least 1) the capacity to communicate the directives to the relevant addressees, and 2) the fact that the (existence of the) directives and their content can be identified without resorting to a balancing of reasons. We should add a third one to these capacities: 3) the capacity to issue directives or utter prescriptions.

(3) Regarding the "type of utterances", there is a multiplicity of kinds of statements with which a subject seeks to guide others' behaviour. What all these kinds have in common is the fact that they are attempts by the utterer to make the addressee act or not act in a certain way.[26]

---

[22] protected reason for action is a systematic conjunction of a first-order reason to φ or not- φ, and a second-order reason to exclude all the reasons against the action contained in the first-order reason; see Joseph Raz, *Practical Reason and Norms*, 2 ed (London: Hutchinson, 1999): 191. A brief *caveat* here: this definition corresponds to practical authority; the type of reasons a theoretical or epistemic authority offers (or pretends to offer) are reasons for belief, not reasons for action. We will briefly come back to this in point II.3 below.

[23] "Subject" here is used in the widest possible sense, encompassing physical people, legal institutions (like Parliaments or Assemblies) and other kinds of entities (like legal systems, artificial intelligence), among others.

[24] In what follows, we will leave aside the debate in terms of reasons for action.

[25] The normative conditions are related to the epistemic capacity to determine what the balance of dependent reasons determines in a concrete case, thus the capacity to know or access moral reasons for action and the capacity evaluate and act in accordance with known moral reasons. In the framework of his Service Conception of legitimate authority, these capacities manifest in the verification of the Normal Justification Condition and the Independence Condition and conduces to the Pre-emption Thesis. However, as our inquiry here is not normative, we will leave this aside for now.

[26] See Cleo Condoravdi & Sven Lauer, "Imperatives: Meaning and illocutionary force", in Christopher Piñón (ed), *Empirical issues in syntax and semantics* 9 (2012): 37–58; John Searle, "A taxonomy of illocutionary acts",

When discussing authority, it is commonplace to consider that these statements are not advice or suggestions but instead directives, norms or commands: imperative statements *requiring* an addressee to act or not to act in a certain way, that is, establishing that the addressee has a duty to act or not to act without any consideration of their will – a duty that does not relate to the content of the proposition but with the will of the utterer. Imperatives, or better still, the utterance of imperatives can be seen as expressing some content, conveying some volition of the speaker, and (pretending to) acting as an inducement for the addressee to conform with the content.[27]

(4) Regarding the "intention", this element seems linked with the analysis of the "type of utterance" in two ways: regarding the exact kind of object they are and regarding the origin of their meaning and force. We will further develop the former in the next points. Regarding the latter, albeit discussion on this point is still ongoing,[28] imperatives have been generally considered within legal theory to be either depending on a volition, desire or will to exist as such and/or to have normative or compulsory force.[29] We will come back to this in the next sections.

(5) Regarding the "norm-dependence", albeit it can be arguably said that all authority – *qua* normative power – is ultimately dependent on a norm, the status of "authority" in law is always dependent on a power-conferring norm belonging to a specific legal system. In this sense, in law, the status of "authority" related to any subject is bestowed to them by a norm establishing, at the very least, the type of power bestowed, the conditions (procedural, substantive) under which that power is to be exercised, and the effects or outputs of its exercise. Following this idea, then, the criterion for allocation of this power to produce certain effects or outputs – such as legal texts or legal norms – by means of certain acts under certain conditions within a certain legal system, and thus to deem a subject an authority (of a certain kind), is completely dependent on that same system.

Let us know circle back to (1) and the question left open there. As we have seen, a bare and basic structural way to conceive an authoritative relationship can be taken as having three relevant parts: an author, an output, and an addressee. An author that produces some output,

---

in Keith Gunderson (ed), *Language, mind, and knowledge* (Minnesota, University of Minnesota Press, 1975): 344-369.

[27] See Condoravdi & Lauer, "Imperatives: Meaning and illocutionary force".

[28] For a great overview, see Alessio Sardo, "The Dark Side of Imperatives", in Francesca Poggi & Alessandro Capone (eds), *Pragmatics and Law: Practical and Theoretical Perspectives* (Cham: Springer Verlag, 2017): 243-271.

[29] See Jeremy Bentham, *The Limits of Jurisprudence Defined* (New York: Columbia University Press, 1945); John Austin, *The Province of Jurisprudence Determined* (Cambridge: Cambridge University Press, 1995). *Contra,* see Olivecrona, *Law as fact,* 2 ed.

This view is, for example, partially in agreement with the so-called standard view of imperatives in speech acts theory but differs in an important point. The agreement is that any (happy) utterance of an imperative conveys both a representational content (action to be done by a certain subject) and the wish that the addressee performs that action, so it assumes a volitional content associated to the imperative. The difference is that it is not that volitional content what gives the statement its illocutionary force: for this view, "*by convention,* certain sentences get to be associated with a certain illocutionary force, through some illocutionary force indicating devices" (Sardo, "The Dark Side of Imperatives", p. 259). This is not to say that all imperatives create obligations for the addresses "by virtue of linguistic convention" because "imperatives are also used with a weaker directive force in requests, pleas, warnings, etc." (Condoravdi & Lauer, "Imperatives: Meaning and illocutionary force", p. 38).

and an addressee that receives (and needs to decide about) this output. The characteristics of each of these parts and their articulation naturally depend on an array of theoretical commitments that could vary – and, in some cases, also practical commitments to some ideals.[30] Two of the main questions to be answered are the following: does an "author" need any factual intention in the production of the output for it to "count as law", or would it be sufficient for it to be presupposed in some context, if any? Does the addressee have any role in the conformation of the characteristics of the output through their interpretation of it, or is it to be considered a mere passive part?

*II.2. The Author: Law as an intentional creation*

Statutes, contracts, and other sources of law are more often than not the product of the activity of a determinate and normatively authorised author who issued or uttered them with the intention of regulating human behaviour. It is considered that most of contemporary law is the product of intentional creation.[31] This intentional character of law has recently developed as a widely held view that law is an artefact.[32] On the most basic dictionary level, artefacts are objects made by human beings. However, the philosophical definition of artefacts in its classical form emphasises that the objects that we characterise as artefacts must be made intentionally, with the aim of achieving some purpose.

For Risto Hilpinen, the very nature of artefacts is an authorial intention.[33] Namely, the term 'author' – or 'group of authors' – refers to persons who cause intentional or unintentional products to exist or events to occur. When the author intentionally creates a product, we call their production authorship and their product, an artefact.[34] Intentional creation by a determinate author is, therefore, a condition for something to be an artefact. Other authors emphasise that for an artefact to exist, the proper function of a thing must be one of the functions that the author intended.[35] Randal Dipert is clear in that artefacts are to be understood by reference to the intentions of the creators of the artefacts.[36] In fact, most philosophical

[30] For example, one of them is the well-known ideal of democracy: roughly, that political authority ultimately derives its power directly or indirectly from the people. Consequently, the institutional design of any State shall reflect this circumstance by assigning a prominent role in law-creation to any institution that represents, directly or indirectly, the people. This was the argument behind the Enlightenment ideal of the Parliament (representatives of the people, expressing the will of the people) as supreme authority fully concentrating the law-creation function and the courts (not representatives of the people) as completely subordinate, "mouth of the law", mechanical law-appliers. This ideal is manifests itself in contemporary debates such as the one regarding the legitimacy and opportunity of judicial constitutional review ("*Should a group of non-representatives of the people decide on the legitimacy of decisions of a group of representatives of the people?*").

[31] See e.g. Joseph Raz, "Intention in Interpretation", in Joseph Raz, *Between Authority and Interpretation* (Oxford: Oxford University Press, 2009): 265-298.

[32] See e.g. Luka Burazin, Kenneth Himma, & Corrado Roversi (eds), *Law as an artifact* (Oxford: Oxford University Press, 2018).

[33] See Risto Hilpinen, "Authors and Artifacts", *Proceedings of the Aristotelian Society* New Series 93 (1993): 155-178.

[34] Hilpinen, "Authors and Artifacts", p. 156.

[35] See e.g. Lynne Rudder Baker, "The ontology of artifacts", *Philosophical Explorations* 7 (2006): 99-111, p. 102.

[36] See Randall R. Dipert, *Artifacts, Art Works, and Agency* (Philadelphia: Temple University Press, 1993).

literature uses "artefact" to mean only those things that are intentionally created for some purpose.[37]

The issue has split the literature on artefacts in two camps. On the one hand, the intentionalist camp argues that human intentions determine the function of the artefact and with it, the artefactual nature of a thing; on the other hand, the reproductive or non-intentionalist camp claims that the artefactual nature of the thing is determined by the history and context of the use of the thing. However, both accounts argue that intention is necessary for the determination of the artefactual character of an object, differing only in whether it is sufficient or not.[38]

Consequently, many legal philosophers have followed this insight to formulate accounts of the functions of law, speculating about the fundamental uses of law as an artefact. After all, views as different as Kelsen's, Austin's, Finnis' and Raz's converge in holding that law is, with the eventual exception of customs and some moral norms, created by "purposive actions of officials".[39] Some have explicitly connected the artefactual character of law with the idea that intention is paramount when discussing the artefactual character of law[40] or when explaining law in itself.[41] Andrei Marmor assumes that artefacts are objects created by humans and used for a typical purpose that might be a part of the authorial intention or not.[42] Being created by humans and used for the intended purpose, for Marmor, the law is an intangible prefixed compound artefact. It is a compound artefact because it operates within other artefacts (conventions, in his view). It is prefixed because "the law is what various authorities say that it is".[43] Law is not a natural kind because it was created by human beings.[44]

If we assume that institutions are agents who can act intentionally as we often do,[45] it seems to follow that law-creation is impossible without intention. Legislation can connect certain

---

[37] See e.g. Beth Preston, *A Philosophy of Material Culture: Action, Function, and Mind* (New York: Routledge, 2013), p. 5.

[38] See e.g. Beth Preston, "Philosophical Theories of Artifact Function", Anthonie Meijers (ed), *Philosophy of Technology and Engineering Sciences* (North Holland: Elsevier, 2009): 213-233; Kenneth Himma, "The Conceptual Function of Law: Law, Coercion, and Keeping the Peace", in Luka Burazin, Kenneth Himma, & Corrado Roversi (eds), *Law as an artifact* (Oxford: Oxford University Press, 2018): 136-159, p. 142.

[39] See Brian Bix, "Obligations from Artifacts", in Luka Burazin, Kenneth Himma, & Corrado Roversi (eds), *Law as an artifact* (Oxford: Oxford University Press, 2018): 163-176, p. 164. However, Finnis gives ample place in this theoretical framework for customary law when analysing law not in terms of authority but of authoritative rules. He is interested in explaining how authoritative rules arise within a certain community, as the existence of those rules is the fact that allows for coordinate action by the community members. See e.g. John Finnis, *Natural Law and Natural Rights*, 2 ed (Oxford: Oxford University Press, 2011): 128, 238.

[40] See Veronica Rodriguez-Blanco, "Processes and Artifacts: The Principles Are in the Author Herself Luka Burazin, Kenneth Himma, & Corrado Roversi (eds), *Law as an artifact* (Oxford: Oxford University Press, 2018): 192-214.

[41] See Kenneth Ehrenberg, *The Functions of Law* (Oxford: Oxford University Press, 2016).

[42] Andrei Marmor, "Law, Fiction, and Reality", in Luka Burazin, Kenneth Himma, & Corrado Roversi (eds), *Law as an artifact* (Oxford: Oxford University Press, 2018): 44-60, p. 45.

[43] Marmor, "Law, Fiction, and Reality", p. 57.

[44] See Frederick Schauer, "Law as Malleable Artifact", in Luka Burazin, Kenneth Himma, & Corrado Roversi (eds), *Law as an artifact* (Oxford: Oxford University Press, 2018): 29-43, p. 29. *Contra*, considering law as a natural kind in the social domain, see e.g. Muhammad Ali Khalidi & Liam Murphy, "Disagreement about the kind law", *Jurisprudence* 12: 1-16.

[45] See Joseph Raz, "Intention in Interpretation", in Joseph Raz, *Between Authority and Interpretation* (Oxford: Oxford University Press, 2009): 265-298, p. 280. It is however doubtful whether this claim is true – and if true,

legal consequences with an event only if its creator (in the broadest possible sense) intends to do so. An agent who is a legal authority creates law only insofar as she performs an act that expresses her intention to create this very law (or to make this particular text count as law from that moment on). A statement on the content of law becomes law when this intention of the intentional agent is expressed in a speech-act.[46] Hence, intention to legislate seems constitutive of the law-creating act, in the sense that law cannot be created absent an intention on the part of the author.[47]

### *II.3. The Reader: Understanding authorial utterances*

#### II.3.1. Intentional creation and intentional interpretation

Interestingly enough, intention is thought of not only as a condition for law-creation, but also as a condition for law-interpretation. In discussing standard accounts of authority, Heidi Hurd claims that all those accounts are closely connected to intentionalist theories of interpretation.[48] While this connection is also noticeable in instances of inspirational and influential authority, it has been explicitly theorised in accounts of practical authority based on the work of Joseph Raz. In his discussions about interpretation and legal theory, Andrei Marmor writes that the acknowledgement of practical authority to another makes it "sensible" to "take the authority's intentions into account when his directives require interpretation".[49] Furthermore, in intentionalist theories of legal interpretation, it is often argued that intention is a regulative rule directing interpreters to seek the will of reasoning of the legislator, but also a constitutive rule without which the cognitive activity of determining the meaning of a legal text could not be considered interpretation at all.[50]

And indeed, according to many authors in legal theory, interpretation is closely connected to intentions.[51] For one, it is claimed that as far as the law is a product of intentional creation, intention is necessary to understand the content of a normative utterance. For Raz, this means

---

to which extent and in which sense it would be true. It seems clear that multimember bodies do not have the same kind of intention as individual agents do. The identification of intention in these cases doesn't rely on a mental state, but on a procedure that has been followed in order to reach a legislative decision. We will come back to this in point IV.2 below.

[46] Raz, "Intention in Interpretation", p. 283.

[47] It is possible to link this point with the point regarding imperatives that was made in two previous sections. It was said that, when uttering an imperative or issuing a directive, the "author" seems to have a conduct-guiding intention. However, this might be too simplistic, especially when considering law-creation within established legal systems. Assuming we accept that intention is involved, it is possible to claim that there are several intentions involved in law-creation: a general intention to make 'law' in order to conduct-guiding; a particular intention to make a certain object O count as 'law' from the moment of the act on; a specific intention to make this object O count as 'law' in order to bring about behaviour Z by the addressees; and so on. We will come back to this in point II.3.2 below.

[48] See Heidi Hurd, "Interpreting Authorities", in Andrei Marmor (ed), *Law and Interpretation. Essays in Legal Philosophy* (Oxford: Oxford University Press, 1997): 405-432.

[49] Andrei Marmor, *Interpretation and Legal Theory* (Oxford: Hart Publishing, 2005), p. 178.

[50] See Bojan Spaić, "Normativity of Basic Rules of Legal Interpretation", in Kenneth Himma, Miodrag Jovanović & Bojan Spaić (eds), *Unpacking Normativity* (Oxford: Hart Publishing, 2018): 157-175.

[51] This is not the case with all legal theorists, nor the ones writing about the artefactual character of law. In his book *The Functions of Law*, Kenneth Ehrenberg is quick to point out that his analysis of the artefactual character of law is "does not imply anything about the relevance of drafters' intentions to the interpretation and application of law once enacted". See Ehrenberg, *The Functions of Law*, p. 1.

that the norm subjects cannot interpret legislation if they do not seek the intention of its creators.[52] According to him, all attempts at understanding the law should reflect the lawmaker's intentions as long as the law is derived from deliberate law-making.[53] Raz formulates his claim as a conceptual thesis, insisting that the search for intention *constitutes* interpretation. However, the thesis contains some evident and strong normative elements. If intention is the foundation of law creation, the interpreter should, in fact, look for this intention and give it preference over any other interpretative approach.

Other authors theorise intention in interpretation without deriving any normative consequences from their conceptual arguments. Stanley Fish, the literary critic turned legal philosopher, has argued in much of his work that interpretation is impossible without intention. Contrary to Raz, intention for Fish is not just an approach to interpretation among others. The presupposition of intention is itself the condition of the possibility to understand any designed or purposively produced object.[54] Yet, this fact tells us nothing about interpreting legal – or any other kind of – texts or materials. Intentionalism, in this sense, is simply a theoretical answer to the question about the intelligibility of anything produced by a "purposive agent," a fact that separates "mere objects" from "messages." Larry Alexander and Saikrishna Prakash are even clearer about the matter: there is no meaning in the lack of an author, and there is no meaning in the lack of an intended meaning.[55] Fish sums it up like this:

> Absent the presumption that the shapes you have encountered—c, a, t, black marks on a white background— are designed, are purposively produced, there is nothing to be done with them in the way of meaning. And if you do something in the way of meaning with them, it is because you are already assuming design, assuming intention, whether self-consciously or not. That is, you have no choice but to do so.[56]

### II.3.2. Kinds of intention

This brings us to our second crucial question. We argued that much of legal philosophy claims that intention is a condition of the possibility of creating law. Given that our accounts of (practical) authority are premised on the possibility that authority has an intention that can be expressed in the form of legal texts, our accounts of legal interpretation often rely on the idea that there is some kind of intention behind every legal text. They also often rely on the assumption that law is, at the core, a communicative phenomenon. An example of the latter are accounts that understand the legal phenomenon using the theory of communication formulated by Paul Grice, which gives paramount importance to the intention of the law-creating authority in determining the meaning of the legal text.[57] Within this approach, a parallel can be established between ordinary conversation and what happens in the legal phenomenon: a legal authority (speaker) communicates a message (directive) to the citizens and other legal authorities (listeners) against a background of a mutually known context between the speaker and listener(s) and where all involved parties have a cooperative predisposition in following certain communication principles.[58] As in ordinary conversation, the communicative intention of this legal authority would be both constitutive of the meaning of the message and the central criterion of correction for decoding – or interpretation – of the message by citizens and other legal authorities.[59]

---

[52] Raz, "Intention in Interpretation", p. 267.

[53] Raz, "Intention in Interpretation", p. 275.

[54] Stanley Fish, "Intention Is All There Is", *Cardozo Law Review* 29 (2008): 1109-1146, p. 1112.

[55] See Larry Alexander & Saikrishna Prakash, "'Is That English You're Speaking?' Why Intention Free Interpretation is an Impossibility", *San Diego Law Review* 41 (2004): 968-996.

[56] Fish, "Intention Is All There Is", p. 1111-2.

[57] For a full development of Grice's theory, see Paul Grice, *Studies in the Way of Words* (Cambridge, Mass.: Harvard University Press, 1989). For an example of an account that claims that the Gricean model is applicable to the legal phenomenon, especially to legal interpretation, see Richard Ekins, *The Nature of Legislative Intent* (Oxford: Oxford University Press, 2012).

Needless to say, there is a vast amount of literature on the nature of intentions. Multiple definitions have been offered: among many others, intentions are conceived as practical attitudes that play a crucial role in plans that agents commit themselves or others to[60] and as mental states that act as reasons for undertaking an action.[61] For our present purposes, however, it is much more important to differentiate between kinds of intention – an important move often lacking in legal philosophy literature. Departing from the lack of discussions about intentions related to the law-creation activity, Marcin Matczak distinguishes between three kinds of intentions. The first kind is the legislator's intention regarding the meaning of the words used in a legal text. According to Marczak, the second kind of intention is the intention to make law, which gives legal force to words. Finally, there is an intention on the part of the legislator regarding the changes that the legal text, endowed with legal force, will bring about in reality.[62] Francesca Poggi distinguishes two different objects of legislative intention and, thus, two distinct shapes: (1) the intention of approving a specific text X and/or (2) the intention of approving a specific text X that expresses a particular norm N.[63] Finally, even within the artefact literature, intentions are sorted into at least two categories. For Hilpinen, artefacts depend on an intention, the content of which is a *description of a type of object* that an author creates.[64] For Baker, intention is the ability of the artefact to perform a specific "proper" function, regardless of the attitudes of the author of the artefact.[65]

---

[58] Otherwise, according to Grice, the communication would fail. See Grice, *Studies in the Way of Words,* p. 26.

[59] Some recent applications of Gricean semantics to law include e.g. Paolo Sandro, *The Making of Constitutional Democracy* (Oxford: Hart Publishing, 2022); Izabela Skoczeń, *Implicatures within Legal Language* (Cham: Springer, 2019).

[60] Michael Bratman, *Intention, Plans, and Practical Reason* (Cambridge, MA: Harvard University Press, 1987), p. 18-20.

[61] See G.E.M. Anscombe, *Intention*, 2 ed (Oxford: Blackwell, 1963), p. 11-15. For an overview of philosophical conceptions of intention see: Kieran Setiya, "Intention", *The Stanford Encyclopedia of Philosophy*. Available at https://plato.stanford.edu/archives/win2022/entries/intention

[62] Marcin Matczak, "Three Kinds of Intention in Lawmaking", *Law and Philosophy* 36 (2017): 651-674, pp. 652-3.

[63] Francesca Poggi, *Il modello conversazionale. Sulla differenza tra comprensione ordinaria e interpretazione giuridica* (Pisa: Edizioni ETS, 2020), p. 232. She also differentiates between four levels of legislative intention, adapting from four levels of Gricean speaker's intention, for the purposes of criticising this comparison: "(i) the intention to utter (say, write) something; (ii) the intention to utter (say, write) what is said (x) and not something else; (iii) the intention to mean (implicate, communicate) something (other/different) by uttering x; and (iv) the intention to mean (implicate, communicate) S and nothing else by uttering x" (Poggi, *Il modello conversazionale*, parag. 27). Nevertheless, these levels are not to be considered as different kinds of intention: they are all part of the communicative intention.

[64] Hilpinen, "Authors and Artifacts", p. 158.

[65] Baker, "The ontology of artifacts", p. 102.

At least three kinds of intention seem relevant for our purposes. First, there is the intention to produce an object type – for example, the intention to produce a rulebook that regulates the behaviour of persons. We can call this general level intention *type intention*. Second, we can have the intention of producing a token of a type – for example, the intention to create a rulebook with determinate textual content on the behaviour of students in the student canteen. We can call this level of intention *token intention*. These first two kinds roughly correspond to the kinds of intention identified within the literature on artefacts. Finally, we could speak about the intention of the rulebook to cause a certain kind of behaviour. Namely, when a legislator enacts a legal text, they most probably have a representation of at least some desired behaviours that the rules derived from the text shall elicit in their addressees. We can call this *causal intention*.

Now, it is not difficult to see that one kind of intention does not necessarily entail the other. A legislator might have a type intention to create law in order to, for example, bring her legal system in line with the legal rules of an international organisation, without having any intention to influence behaviour. Or it might be that there is an intention to change behaviour without any intention of producing a statute or even law. Finally, we could intend to create a token of a legal act without the intention to make law or change behaviour. The act might be an example for students to use in Introduction to Law classes.[66]

*II.4. Summing Up*

Up to this point, the relevant conclusions from the revisitation of literature are that:

1. The possibility of an agent having a mental state that we call intention is, it seems, often postulated as the condition of the possibility of law-creation by an agent.
2. The existence of an intention behind the law is the condition of the possibility of law-interpreting.

Until a couple of years back, the entire debate might have seemed scholastic since most of the law seemed to be, in fact, intentionally created, and it was hardly imaginable for a legal text to be authored without any intention of creating legal texts. It would be followed by arguments regarding alternative approaches to interpretation, and at most, include attempts at experimental philosophy aimed at determining the attitudes of legislators and legal interpreters regarding intention in law-creation and law-interpretation. Neither law-creation nor law-

[66] Several things can be discussed here regarding the possible relations between these kinds of intentions and the subject-matter of each one of them. Regarding the former, on the one hand, there might be a relation of implication between type and token intention, in the sense that it is possible to conceive a case where there is type intention and no token intention; on the other hand, it is *prima facie* difficult to conceive a case where there is of token intention and no type intention. Causal intention, for its part, seems to not have any specific relation of this kind neither to type nor to token intention. Regarding the latter, type and token intention seem to have law-creation *stricto sensu* as subject-matter: that is, the act of law-creation itself. The open question is the exact output of this creation: it could, for example, refer to the creation of legal texts and/or the creation of legal norms (or the creation of legal texts that expresses specifically, and uniquely, certain legal norms), as Poggi's taxonomy was referring. Causal intention, for its part, seems to have law-creation *latu sensu* as subject-matter: that is, the producing of certain consequences or changes in reality (by means of law-creation).

We will come back to this discussion in the next sections.

interpretation, for all the talk about legal philosophy possibly (or hopefully) becoming mechanical, were considered activities that might imaginably be delegated to hardware or software, no matter how advanced. However, as we saw at the very beginning of the paper, recent technological developments – particularly the advent of generative artificial intelligence – have completely changed the picture.

## III. INTERLUDE: AGENTS WITHOUT INTENTIONS

### *III.1. How did we get here?*

Artificial intelligence has been a prominent topic in legal science and philosophy of law for the last three decades. It started with a keen interest in reasoning with legal cases, representing legal knowledge, and modelling deontic concepts. The beginning of the new century was marked by an increased interest in modelling legal knowledge as an answer to the ever-growing amount of legal information available in electronic form, along with argumentation schemes. Finally, in the last decade, advanced machine learning techniques have been rising to prominence as the main technology driving the interplay between law and artificial intelligence.[67]

A major development leading to the current explosion of AI applications occurred independently of the legal interest in AI. In 2017, Google engineers and scientists invented the transformer architecture that became the foundation of the current large language models (LLM) of AI. *Transformers* are an architecture for deep learning, computer models for natural language processing introduced in a 2017 paper entitled "Attention is all you need".[68] They "understand" words in sentences and their meaning by "focusing" on words and their contexts. This allows the model to understand language and output meaningful language, allowing breakthroughs like GPT, Claude, LaMA, Gemini and others. The models have found use in legal research by providing summaries of complex legal texts, identifying relevant case laws, statutes, or legal arguments; contract analysis by reviewing contracts and other legal documents, highlighting potential issues; drafting legal documents, analysing case law, and statute law to predict legal outcomes; providing legal assistance with LLM chatbots.

### *III.2. Creating v. Generating legal texts*

Legal scholars have been cautiously probing LLMs since their commercial adoption in early 2023. The existing studies demonstrate impressive capabilities in various domains of the legal profession. They have tested the abilities of the most capable of the models to enrol in a law school with LSAT tests[69], its ability to pass law school exams[70], its success in interpreting statutes[71], the possibility of teaching a model to think like a lawyer[72], the ability of ChatGPT to differentiate between legal reasoning and other forms of reasoning[73], and even the possibility of latest generation models to pass the bar exam[74]. GPT-4 did particularly well on exams requiring understanding texts and their interpretation and writing short prose. One of the most detailed studies concluded that "large language models can meet the standard applied to human lawyers in nearly all jurisdictions in the United States by tackling complex tasks requiring deep legal knowledge, reading comprehension, and writing ability."[75]

[67] See Trevor Bench-Capon, "Thirty years of Artificial Intelligence and Law: Editor's Introduction", *Artificial Intelligence and Law* 30 (2022): 475-479.

[68] Ashish Vaswani *et al*, "Attention is all you need", in I. Guyon *et al* (eds), *Advances in neural information processing systems* 30 (California: Curran Associates, Inc., 2018): 5999-6009.

[69] See e.g. OpenAI *et al*, "GPT-4 Technical Report", *arXiv* (2023), p. 5. Available at https://doi.org/10.48550/arXiv.2303.08774

[70] See e.g. Jonathan Choi *et al*, "ChatGPT goes to Law School", *Journal of Legal Education* 71 (2022): 387-400.

In many ways, considering the technology behind LLMs, the results are to be expected. Namely, law is closely related to language, and the language processing capabilities of GPT-4 are impressive. Most of the caution related to large language models until the beginning of 2023 was mainly related to the fact that the technology could safely be disregarded as a serious contender in the domain of law. Nowadays, most concern is associated with the possibility of aiding lawyers to such a degree that lawyers mostly worry about the possibility of machines stepping in place of humans in some legal professions. For better or worse, law firms have been running to include fine-tuned versions of existing LLMs in their workflows. With impressive legal drafting capabilities, fine-tuned models can produce impressive legal documents, especially in the domain of private law.

The capabilities of commercially available LLMs are currently limited in the domain of drafting law. A basic prompt, "*Act as a legislator. Draft a statute prohibiting the carrying of guns in public places",* produces the following output by February 27th, 2024, version of ChatGPT based on GPT-4:

**[Your State/Country] Public Safety Act on the Prohibition of Carrying Firearms in Public Places Section 1: Short Title**
This Act may be cited as the "Public Safety and Firearms Control Act of [Year]."
**Section 2: Definitions**
For the purposes of this Act:

a. "Firearm" means any device, regardless of its name, that is designed to expel a projectile by the action of an explosive, a compressed gas, or any other mechanism.
b. "Public place" means any area that is open to the public or a substantial group of persons, including but not limited to, buildings, roads, parks, schools, places of worship, shopping centers, malls, sports arenas, restaurants, theaters, and public transportation facilities.
c. "Carrying" includes the act of holding a firearm on one's person, or in a container in such a way that the firearm is readily accessible for immediate use.

**Section 3: Prohibition**

a. No person shall carry a firearm, openly or concealed, in any public place within [Your State/Country].

---

[71] See e.g. Andrew Blair-Stanek *et al*, "Can GPT-3 Perform Statutory Reasoning?", *ICAIL '23*: *Proceedings of the Nineteenth International Conference on Artificial Intelligence and Law* (New York: Association for Computing Machinery, 2023): 22-31.

[72] See e.g. Fangyi Yu *et al*, "Legal Prompting: Teaching a Language Model to Think Like a Lawyer", *arXiv* (2022). Available at https://doi.org/10.48550/arXiv.2212.01326

[73] See e.g. Bojan Spaić & Miodrag Jovanović, "Artificial Reason and Artificial Inteligence: The Legal Reasoning Capabilities of GPT-4", *Annals of the Faculty of Law in Belgrade* 72 (2024): 383-422.

[74] Katz *et al*, "GPT-4 Passes the Bar Exam", p. 10.

[75] Katz *et al*, "GPT-4 Passes the Bar Exam", p. 10.

b. This prohibition does not apply to:
   - i Authorized law enforcement officers in the performance of their official duties.
   - ii Members of the armed forces of [Your State/Country] or of the federal government, when on duty and required to carry firearms as part of their official equipment.
   - iii Licensed security personnel while engaged in official duties.
   - iv Individuals specifically authorized by a written exemption from the [Relevant Authority], under conditions that ensure public safety.

**Section 4: Exemptions**

The [Relevant Authority] may grant exemptions to this prohibition under strict criteria that consider the necessity of carrying firearms for specific, lawful purposes such as hunting, sport shooting, and other activities deemed not to pose a public safety risk.

**Section 5: Penalties**

a. Any person found in violation of this Act shall be subject to penalties including, but not limited to, fines, imprisonment, or both, as determined by the judiciary of [Your State/Country].

b. The severity of penalties will reflect the nature of the violation, any prior offenses, and the circumstances surrounding the violation.

**Section 6: Implementation and Enforcement**

a. The [Relevant State/Country Department] shall be responsible for the implementation and enforcement of this Act, including the development of guidelines for the lawful carrying of firearms for specific purposes as outlined in Section 4.

b. This department shall also provide public education on the requirements of this Act and the processes for applying for exemptions.

**Section 7: Severability**

If any provision of this Act, or its application to any person or circumstance, is found to be unconstitutional or otherwise invalid by a court of competent jurisdiction, the remainder of the Act and its application to other persons or circumstances shall not be affected.

**Section 8: Effective Date**

This Act shall take effect [number of days] days after its enactment.

The short version of the act that GPT4 can produce based on a simple prompt can, of course, be expanded by the model, but the result is not dramatically different. In lack of legal fine-tuning, the results are overly generic and heavily dependent on the abilities of the human prompter. This aligns with the research on prompting and the limits of generative AI models that have recently emerged.[76] Based on those examples and the outputs that are generated with well-crafted and detailed prompts, even if we are not inclined to claim that AI can in fact, create legal texts, in the sense of texts that could count as law, it seems indisputable that it can generate legal-like texts hardly distinguishable from human created legal texts. This is not only noted by the cited studies and literature but also suggested by real cases such as the Brazilian one – where everyone was none the wiser about the origins of the bill they voted on until the councilman decided, for some reason, to disclose the information.

## IV. LEGAL AUTHOR(ITY): SECOND TAKE

[76] Yu *et al*, “Legal Prompting: Teaching a Language Model to Think Like a Lawyer”.

Arguably, generative artificial intelligence can be developed further to be able to significantly participate in drafting law either by fine-tuning or other successful techniques like retrieval augmented generation (RAG). It is, however, dubious whether the development of the current models will ever lead to the development of intelligence, reasoning, cognitive abilities, intentions, or mental states. When asked about the presence of intention in his outputs, ChatGPT replies, "I don't have intentions, emotions, or consciousness". By now, we know that LLMs do not say what they "think"[77] and that every answer should be taken with a grain of salt if we want to infer something about the inner workings of the transformer architecture.[78] Namely, the companies and researchers behind the models have been so enthusiastic about their abilities and somewhat frustrated by their lack of transparency that they shared the tendency to anthropomorphise LLMs. OpenAI researchers noted in the GPT-4 technical paper that LLMs can "create and act on long-term plans" displaying even power-seeking behaviour that is becoming increasingly "agentic".[79] Research routinely talks about reasoning[80], analogical reasoning[81], and abstraction capabilities[82], but for all the talk about reasoning as identifiable in the outputs of LLMs, there is no indication of anything resembling internal life. Moreover, the increase in size of LLMs seems to make them less transparent, and their outputs further distanced from their inner workings.[83] What is more, there are indications that the success in displaying reasonable-sounding outputs can at least partially be explained by the term frequencies in datasets.[84]

Be this as it may, even emergent behaviour is only displayed, outputted, and outward-facing, and there is no evidence of LLM's internal life to speak of. Even the studies of consciousness in its most analytically striped form, arguing that there are indicator properties of consciousness, conclude that "no current AI systems are conscious".[85] In this sense, we can confidently claim that nothing present in current-generation AI is even remotely similar to the forms of intelligence known to us. If anything, the intelligence that the outputs of LLMs exhibit is so foreign that, at best, it can be considered an alien form of intelligence that we have yet to understand.[86] Consequently, it is possible to conclude that generative artificial intelligence models do not possess mental states, dispositions, or attitudes that constitute intentions in the sense described in the previous parts of the paper.

---

[77] Miles Turpin *et al*, "Language Models Don't Always Say What They Think: Unfaithful Explanations in Chain-of-Thought Prompting", *arXiv* (2023). Available at https://doi.org/10.48550/arXiv.2305.04388

[78] Roman Yampolskiy, "Unexplainability and Incomprehensibility of AI", *Journal of Artificial Intelligence and Consciousness* 7 (2020): 277-291.

[79] OpenAI *et al*, "GPT-4 Technical Report", pp. 53-4. The competitive nature of LLM development has taken its toll, and the more recent technical papers released upon deployment of new models do not contain many references to human-like abilities of LLMs. See Gemini Team Google *et al*, "Gemini 1.5: Unlocking multimodal understanding across millions of tokens of context", arXiv (2024). Available at https://doi.org/10.48550/arXiv.2403.05530

[80] Jie Huang *et al*, "Towards Reasoning in Large Language Models: A Survey", *arXiv* (2022). Available at https://doi.org/10.48550/arXiv.2212.10403

[81] Taylor Webb, Keith Holyoak & Hongjing Lu, "Emergent analogical reasoning in large language models", Nature *Human Behaviour* 7 (2023): 1526-1541. There are of course outliers, arguing that "artificial intelligence is stupid". See J. Mark Bishop, "Artificial Intelligence Is Stupid and Causal Reasoning Will Not Fix It", *Frontiers in Psychology* 11 (2020): 1-18.

[82] Melanie Mitchell, Alessandro Palmarini & Arseny Moskvichev, "Comparing Humans, GPT-4, and GPT- 4V On Abstraction and Reasoning Tasks", *Proceedings of the LLM-CP Workshop, AAAI 2024*. Available at https://doi.org/10.48550/arXiv.2311.09247

[83] Tamera Lanham *et al*, "Measuring faithfulness in chain-of-thought reasoning", *arXiv* (2023). Available at https://doi.org/10.48550/arXiv.2307.13702

[84] Yasaman Razeghi *et al*, "Impact of pretraining term frequencies on few-shot numerical reasoning", in Yoav Goldberg, Zornitsa Kozareva, Yue Zhang (eds), *Findings of the Association for Computational Linguistics: EMNLP 2022* (Association for Computational Linguistics): 840-854.

[85] Patrick Butlin *et al*, "Consciousness in Artificial Intelligence: Insights from the Science of Consciousness", *arXiv* (2023), p. 1. Available at https://doi.org/10.48550/arXiv.2308.08708

Still, generative AI models are agents despite lacking mental states or intelligence.[87] They can output relatively autonomous reasonings and even actions if certain preconditions are met. Undoubtedly, the outputs of the most advanced LLMs we had at the time of writing this paper can, for all intents and purposes, be compared with the outputs of a human law-creator. What is more, we are yet to find the technical limitations of LLM that would convince the community that the models are, in principle, unable to do the tasks of law-creation and law-interpretation because of some software limitations or hardware bottlenecks.

However, according to the positions we expounded on in the previous sections, the outputs of ChatGPT cannot be considered law because they do not meet the requirements for texts to be counted as law (see *II.4*). Namely, in lack of anything that we might reasonably call an intention in ChatGPT, we cannot say that the textual outputs resembling legal texts produced by it can count as law. Furthermore, according to the same philosophical reasoning, if there is no intention to identify behind a text produced by GPT, a determination of the meaning of the outputs of the LLMs cannot be considered interpretation (see *II.3.1*).

These conclusions seem counterintuitive. For one, we already know of instances in which real-life legislators used LLMs to draft entire bills and at least one instance where one of those bills has passed and is now part of a legal system. Human interpreters will be faced with the task of interpreting those texts. In our opinion, there are two possible ways out of this: either (1) the claim that intention is a condition of the possibility for law-creation and law-interpretation is true and human intention should be in some way ascribed to the outputs of LLMs, or (2) the claim that intention is a condition of the possibility for law-creation and law-interpretation is false. Let us explore these two possibilities.

### *IV.1. One: His Master's Intentions*[88]

---

Jacob Browning and Yann LeCun argued, in 2022, that "[a] system trained on language alone will never approximate human intelligence, even if trained from now until the heat death of the universe." See Jacob Browning & Yann Lecun, "AI And the Limits of Language", *Noema* (August 23, 2022). Available at https://www.noemamag.com/ai-and-the-limits-of-language/ However, the authors quoted argue that there are no technical limitations to LLMs developing consciousness; see e.g. Butlin *et al*, "Consciousness in Artificial Intelligence: Insights from the Science of Consciousness", p. 1. These two clams are not necessarily incompatible since AI development is pushing towards multimodality, entailing the extension of their capabilities to audio, video, and images.

[86] "*It is not intelligence as we know or understand it*", Mr Spock would say to a puzzled reader.

[87] Luciano Floridi, "AI as Agency Without Intelligence: on ChatGPT, Large Language Models, and Other Generative Models", *Philosophy & Technology* 36 (2023): 36-15.

[88] A nod to Stanislav Lem's novel *His Master's Voice* (1968), a first contact story detailing the attempts to decode an extra-terrestrial message.

If law entails authorship, authorship necessarily entails intention, and ChatGPT doesn't have intention, there is a possibility that intention could somehow be "borrowed" from human beings that operate the AI. The very possibility and degree of possibility of transferring human intention to artificial agents is premised on the current levels and future levels of technological developments. Namely, it stands to reason that even if there is no intention behind the generation of large language models, the human agent that either operates the model or has created the model acts intentionally in the interaction or creation of the model. If this intention is of the right kind, we could say that all the outputs of LLMs are, in fact, indirectly based on human intention, and unintentional AI agents create that law. An educated guess regarding these issues leads us to postulate three possible scenarios:

**1. AI as an extended legal mind.** In the first scenario, the AI is used as a research tool for drafting legal documents. An AI fine-tuned to do legal research, and especially comparative legal research, would be capable of giving the human legal author valuable data on the regulation in comparative and international settings and about the available scientific research data. Given the way LLMs operate, the data is not just presented to the legal drafter but also interpreted by AI to create a meaningful whole. So, while a human being is the author of the text, AI outputs the material that informs the author and serves as a basis for legal drafting. The data is a valuable input for legislators. In this scenario, the very source of law is just informed by AI and not authored by AI. In this case, the legal author remains the human being operating the artificial intelligence. That settles the question related to the presence of intention in law-creation. After all, artificial intelligence is, in this case, used as a tool to achieve the purpose of law-creation. In a way, AI acts as an extension of the mind of the drafter and does not add much more to the mix of law-creation except for adding another modality in which the human mind is extended by technology.[89]

However, even in this scenario, the matter of intention for the understanding of the law created in concert between a human and a machine needs to be revised. Let us go back to the three relevant kinds of intentions related to law-creation that we have distinguished: type intention, token intention, and casual intention. The human drafter has a type intention to create law, but the token intention can, under certain conditions, be problematic. At least some of the material the author uses here is outputted by AI. In this case, while there might be a token intention behind some of the content of the legal text (the one drafted and/or controlled by the human being), other parts would not have this intention behind them (the ones drafted by AI). Still, we could claim that they were acts of law-creation in virtue of the fact that a human being reviewed the AI-produced texts. At least some parts of the text would not fulfil the requirement of being created by an intentional author and would not qualify as law according to the criteria of authoritative production by an intentional author.

**2. Humans as *causa motrix sive eficiens*.**[90] In the second scenario, the human intervention in the generation of legal texts is limited to the formulation of the prompt(s) that are given to the AI for the purpose of producing a legal text. This possibility is already present in the current state of technology, illustrated by the already well-known Brazilian case. The member of the council that has imputed the prompt was clear in stating that the bill resulting from the prompt was in no way substantially modified by the human prompter. AI produces legal texts based on comparative legal regulation and factual data. In this scenario, there is joint authorship of the legal text, with most of the words written by AI and the human operator acting as a prompter. Both AI and human operator are, thus, law-creators.

[89] Andy Clark & David Chalmers, "The Extended Mind", *Analysis* 58 (1998): 7-19.

[90] An allusion to Aristotle's teaching on four causes from his Physics and Metaphysics. Aristotle claim that the knowledge of a thing is premised on our knowledge of the causes of the thing and argues that there are four causes for things. The first one is matter, the second one is form, the third one is movement or efficiency, and the fourth

The prompt is created with the type intention of law-creating and the factual intention of producing determinate results. The token intention that is supposed to be behind the linguistic content of the statute or any other legal act is lacking, as the content of the act was produced by artificial intelligence. In this scenario, law is produced intentionally by virtue of the fact that a human-type intention and human-causal intention are behind the prompt, leading to the output being an intentional creation. Still, when determining the legal or normative content of the legal act for the purpose of its understanding or application, there is no determinable intention behind the words generated by the AI. In other words, this scenario eliminates the possibility of law-interpretation based on the actual token intention of the legislator or legal author.

**3. Autonomous AI generation (the "Postema scenario").** In the third and final scenario, human prompting is wholly eliminated. Even if, at a technological level, this might not be exactly science fiction, at an intuitive level, it still seems to be a very remote possibility, not warranting extensive discussion. However, Gerald Postema has recently popularised this highly speculative standpoint in his last book, where he discusses the implications of AI for the Rule of Law.[91]

At first glance, it is not clear that the Rule of Law is in any way, shape, or form related to artificial intelligence: (1) If artificial intelligence augments human intelligence, the law would, in principle, still be able to restrict the exercise of arbitrary power by the humans whose intelligence is augmented; (2) If artificial intelligence completely displaces the human activities of law-creating, law-applying, and law-adjudicating, which would presumably be possible with the advent of AI singularity – is in a hypothetical future point in time at which technological growth becomes uncontrollable and irreversible, resulting in unforeseeable consequences for human civilisation – the Rule of Law would be, according to many, the least of our concerns.[92] This is perhaps why the problem of AI and the Rule of Law is posited in somewhat radical

---

one is the end or purpose of the thing. In our discussion motion or efficiency is understood as the human input that starts the process of law generation by an LLM.

[91] See Gerard J. Postema, *Law's Rule*: *The Nature, Value, and Viability of the Rule of Law* (Oxford: Oxford University Press, 2022), ch 14.

[92] Even if some form of Rule of Law could survive in the form of a set of rules, for example as "*Twelve Rules of Rule of Law*", incorporated in AI as either hard ethical limits for law-creation (comparable with Asimov's "*Three Laws of Robotics*") or as efficacy-related and efficiency-obtaining parameters for law-creation. However, in the first case, nothing seems to suggest that human ethical principles would be preserved in this described scenario, and *ex hypothesi* no human prompt or programming would have survived. In the second case, the possibility is based on the assumption that (1) there would be a purpose of behaviour-guiding; and (2) the type of behaviour-guiding would be the one that Postema calls "regulation of conduct", when instead it is plausible to think that the most efficacious and efficient way of behaviour-guiding would be the one that Postema calls "regimentation of conduct".

terms in Postema's book. The question Postema posits, which prompts his discussion of this scenario, is: "*What would we lose if we were to displace the law and have AI regulating our lives?"* It entails an AI that could in some way create, apply, and enforce laws, completely displacing human intervention. At this level of technological advancement, we have to presuppose that human beings would still be trying to interpret law created by an autonomous legal author who is an agent lacking mental states.

This final scenario is the most far-fetched but also the most interesting from the perspective of our discussion. In the present state of technology with AI models lacking mental states and, therefore, lacking intention, this scenario completely eliminates the possibility of intention being behind the outputs counting as legal texts and the possibility of interpreting these legal texts by looking for the authorial intention. In other words: according to the widely held view in legal theory and philosophy, all the outputs produced by AI could not be considered to count as law, nor could they be interpreted as law. In this third scenario, any attempt to borrow or rescue the real intention behind the text would fail.[93]

*IV.2. Two: Author(ity) without intentions*

Instead of going through the trouble of figuring out a way in which AI can borrow intention from a human prompter, we suggest following the lead of theorists who have suggested – or can be interpreted as suggesting – that intention is not a necessary condition for an utterance by (a subject that counts as) an authority to be (to count as) law.

In a 1967 essay, the French literary critic and theorist Roland Barthes advanced the argument that the traditions of empiricism, rationalism and reformation had given birth to the idea of the author as a human person, an individual who produced a body of work. This has led to the idea that an author's complete body of work should be united. In literary criticism, this meant a focus on the person's life, tastes, and passions. At the time of writing the essay, Barthes claimed that the idea of the author was already dented, leading to positions that "show that the whole of the enunciation is an empty process, functioning perfectly without there being any need for it to be filled with the person of the interlocutors".[94]

Barthes argued that the refusal to assign an "ultimate" or "fixed" meaning to the text reveals the possibilities within the text, leading to a focus on the reader of the text as the source of unity or disunity of meaning. With the author out of the way, the meanings of texts are not lost. They are multiplied and rendered transpersonal. Consequently, the presupposition of intention is stripped from any normative consequence, as it might have some ontological but little to no

[93] Of course, this wouldn't prevent us to adopt a purposivist approach that conceives of purposes as the intentions of a perfectly rational or perfectly just legislator, or to use counterfactuals to reach conclusions about the meaning. See Canale, D., & Tuzet, G. (2023). Legislative Intentions and Counterfactuals: Or, What One Can Still Learn from Dworkin's Critique of Legal Positivism. *Ratio Juris*, *36*(1), 26–47. However, artefacts can hardly be based on these and while this might be of help in interpreting text using something that one might connect to the intention of the author, the artefactual nature of the text can hardly be based on our imagination regarding an author.

[94] Roland Barthes, *Image, Music, Text* (Hammersmith: HarperCollins UK, 1977), p. 145.

epistemological value. In other words, Barthes argued, we do not need authors for words to have meaning.[95]

From a very different perspective, Larry Alexander forcefully argued that interpretation without intention is impossible. In his view, meaning attribution is impossible without referencing an author for a plethora of reasons: 1) texts do not declare the language they are written in nor 2) the context in which they find themselves; 3) the presupposition of an author makes marks intelligible, 4) speakers often give alternative meanings to words, 5) all other methods of interpretation (like textualism) must make use of the intentionalist repertoire of interpretative tools like context, avoidance of absurdities etc. Alexander formulated his arguments as a response to the growing appeal of textualism in both the interpretative theory and interpretative practice in the United States, which is still dominant.[96] By and large, they closely follow the idea of Stanley Fish that intention is a precondition of interpretation without necessarily being an interpretative tool. However, Alexander includes a normative element in his discussions by claiming that the search for intention is a method of interpretation that includes more or less precise tools for determining meaning.

Leaving aside the issues that might arise from Alexander's arguments, like the conflation between author and intention and the seeming conflation between the reader's meaning and speaker's meaning, his discussions invite us to consider whether it is possible to develop an alternative position in legal philosophy that can eliminate intention from the epistemology of the creation of law, legal authority, and legal interpretation.

### IV.2.1. Do we really need intention?

In much the same fundamental way Barthes argued against the literary idea of the author, it has been argued that intention is unnecessary for law to be created or interpreted. Three strategies can be reconstructed: (1) rejecting the necessity of intention for artefact-creation (*the unintentional author*); (2) claiming that there is no space nor importance for any "intention" of any legal author in the legal output and its functioning (*the unintentional authority*); and (3) claiming that it is impossible in a legal context to find an "intention" of a legal author as we could do it with an ordinary author in ordinary conversation, so the attention must be put on the other part of the relation (*the intentional reader*).

### IV.2.2. The Unintentional Author

The artefactual character of law enjoys wide agreement in the literature. However, there is not much agreement about what this thesis entails from the perspective of the intentional character of artefacts. Beth Preston argues that most of the literature on artefacts is based on discussing the role of human intention. The position that the artefact functions are determined entirely by individual and collective intention is called intentionalist.[97] The view that we presented in earlier parts of the paper is, however, faced with significant challenges, the main one being there are many instances of products of individual or collective behaviour that result in objects without any intention on the part of the actors.[98] The non-intentionalist views of artefact functions argue that there are indeed some functions that artefacts have historically served (the so-called proper functions), but also those functions that the artefacts are not created to serve but are capable of serving (the system functions). The property functions cannot be entirely dispensed with in a general theory of artefacts. The simple reason for this is that without intentions, we would have to exclude completely novel artefacts (prototypes) from the theory.[99] That is to say, completely novel artefacts are not used by their potential users, and no user intention can determine their artefactual nature.

---

[95] Barthes, *Image, Music, Text*, p. 148.

[96] Alexander & Prakash, "'Is That English You're Speaking?' Why Intention Free Interpretation is an Impossibility", pp. 972-982; Larry Alexander & Emily Sherwin, *Demystifying Legal Reasoning* (Cambridge: Cambridge University Press, 2008), pp. 192-200.

[97] See Preston, "Philosophical Theories of Artifact Function".

Some authors in the law as an artefact literature have taken cues from Preston and argued either for a mixed account of artefact functions or an account that is non-intentionalist. Frederick Schauer contends that the main trait of law as an artefact is its contingency.[100] Corrado Roversi criticises both the purely intentional and the purely functional accounts of the nature of artefacts but eventually sides with Randal Dipert in considering the original authorial intention essential for an account of artefacts.[101] For Luka Burazin, the artefactual character of law is equated with the positivity of law – it means nothing more than the fact that "the law [was] posited by the intentional actions of some human authority".[102] The intentionality required for the creation and existence of a legal system is to be understood not as an individual authorial intention but as a collective recognition or acceptance.[103] Dan Priel argues that individual actions that lead to the creation of something might be intentional actions in the sense that they were undertaken by conscious beings with intentions. Still, they need not be intended to create the thing that arose from their fulfilment. He uses this reasoning to claim that some laws, customs in general and common law in particular, are not artefacts in the strict sense.[104] Giovanni Tuzet argues forcefully that there are significant peculiarities in the artefactual character of law. Namely, law is a strange artefact because it's intellectual as opposed to material, normative as opposed to factual, social as opposed to individual, and interpretative as opposed to intentional.[105] This final characteristic is crucial for our discussion. Namely, law is not created individually but in an interplay of legislators, judges, administration, scholars, and legal subjects. The strangeness of law as an artefact emphasised by Tuzet arises from the following facts: (1) there is not a single authorial intention behind law – it is created in different places, for different purposes and at different times, (2) the purposes contained in the original intention tend to change without any change in the law – as a tool, the law is almost bound to be created for one purpose and use for a different one, (3) the application of law entails more than the identification of the intention that might be attributable to the author of the legal text.[106]

[98] See Preston, *A Philosophy of Material Culture: Action, Function, and Mind*, p. 5.

[99] See Preston, "Philosophical Theories of Artifact Function", p. 227.

[100] See Schauer, "Law as Malleable Artifact", p. 43.

[101] Corrado Roversi, "On the Artifactual—and Natural—Character of Legal Institutions", in Luka Burazin, Kenneth Himma, & Corrado Roversi (eds), *Law as an artifact* (Oxford: Oxford University Press, 2018): 89-111, p. 95.

[102] Luka Burazin, "Legal Systems as Abstract Institutional Artifacts", in Luka Burazin, Kenneth Himma, & Corrado Roversi (eds), *Law as an artifact* (Oxford: Oxford University Press, 2018): 112-135, p. 112.

[103] Burazin, "Legal Systems as Abstract Institutional Artifacts", p. 115.

[104] Priel, "Not All Law Is an Artifact: Jurisprudence Meets the Common Law", p. 252.

[105] Giovanni Tuzet, "A Strange Kind of Artifact", in Luka Burazin, Kenneth Himma, & Corrado Roversi (eds), *Law as an artifact* (Oxford: Oxford University Press, 2018): 217-238, p. 218.

We have seen that there could be a multiplicity of intentions behind law as an artefact. However, even those authors in philosophy of law who have argued against strong conceptions of intention as a basis for discussing artefacts are not ready to give up on intention altogether. The same Tuzet argues that it has both intentional and unintentional functions, but this is quite different from claiming that law as an artefact is independent of the intention that gave rise to it. for all the criticism of the specific intentions behind the law, is careful to argue that "some collective intentionality is necessary to confer upon certain acts the status of legality and the capacity to produce instances of law".[107] When discussing the importance of function for law's understanding, assuming that law is a kind of artefact, Ehrenberg distinguishes between "design function" and "use function" and indicates that the former is usually associated with the notion of "proper function": a characteristic end that it is supposed to achieve and that can explain its existence. However, such an approach "does not account for how an artefact's function might change or how it might differ when seen with the eyes of its creator from the eyes of its later users".[108] Related to law, he further indicates that "we generally expect that it is the intention of a creator that assigns artefacts their functions", even if do not know if this would be the intention of the "creator of the genre or the prototype, or of each token".[109] In this opinion, functions of the law might change over time so "there must be a greater emphasis on what is done with the law at the moment than what was done with the law in the beginning".[110]

One possible conclusive reading is that, for law to be considered an artefact, it is not needed to argue that intention is necessary for something to count as law. It might be enough for the relevant community to treat it as law. This functions both at the artefact-type and the artefact-token level. In the first case, even if it seems difficult to conceive artefact-type without the creator's intention, it is entirely possible: it is enough to consider that the "proper function" of the artefact is not the design function but the use function. Even those who argue that proper function and design function are more or less equivalent recognise that the use function can alter the object's identity over time and thus become its proper function at some point. As such, the ultimate criterion for artefact type is not the design function but the use function – and if that is so, the conditions under which and within which the author has created the object seem, ultimately, irrelevant. In the second case, artefact-token without the creator's intention is also entirely possible: either the intention is ascribed by the users based on some requirements being fulfilled (like parliamentary procedures) or, also here, the intention is completely left aside because the ultimate criterion is the use function.

In conclusion, it is perfectly conceivable to have artefacts without any authorial intention behind them as long as the use of artefacts is such that it attributes and maintains a proper function of the artefact.[111] For the purpose of our discussion, within artefact theory, the outputs generated by artificial intelligence can be considered legal artefacts as long as they are used as law inasmuch as it is needed for their proper function to be the legal regulation of human behaviour.

---

[106] Tuzet, "A Strange Kind of Artifact", pp. 228-30.
[107] Tuzet, "A Strange Kind of Artifact", p. 220.
[108] Ehrenberg, *The Functions of Law*, p. 24.
[109] Ehrenberg, *The Functions of Law*, p. 51.
[110] Ehrenberg, *The Functions of Law*, p. 52.

IV.2.3. The Unintentional Authority

One example of the second strategy, claiming that there is no space nor importance for any “intention” of any legal author in the legal output and its functioning, is Karl Olivecrona’s understanding of legal norms as “independent imperatives”. In this view, legal norms are imperatives that do not depend on the will or intention of any author for their content and strength: they achieve their function with complete independence from it just by having been produced with certain formalities and having received a certain label – such as “law”.

The core of this position is his rejection of what he denominates as “will theories”, that is, theories that conceive law as a set of mandates issued by a supreme or sovereign authority, where the mandates would be an expression or declaration of that authority, and this would provide both unity and binding force to them.[112] Olivecrona’s rejection of the existence (or relevance) of an intention or will of the legislator stems from two main arguments.[113] The first is more general, resulting in dispensing with the idea that any 'imperative' should be equated with 'imperative of conduct', *i.e.* 'imperative' equated with 'command'. The second is, perhaps, more trivial: it is the idea that the 'will' of the legislator is a mere fiction, which is also harmful since it conceals the real mechanism of the operation of law. Regarding the former, Olivecrona argues that part of what leads to the need in law to consider the existence of an authority whose intention or will is relevant in relation to legal norms is the consideration that legal norms are imperatives. In this sense, for Olivecrona, it is evident that these are modes of expression used in a suggestive way to influence people's behaviour or, rather, means to inculcate behaviour in individuals in a categorical way. This consideration, Olivecrona argues, is generally accompanied by the idea that imperatives are fundamentally commands or orders. However, not all imperatives are commands, and it is precisely at this point that the idea of an "imperative without will" appears.

In brief, Olivecrona argues that commands imply a type of personal relationship between two subjects, often in physical proximity (same space, same time), where a non-conditional wish or will is expressed for the addressee of the command to act in a certain way. Its "imperative force" – that is, its suggestiveness – depends mainly on the expression and bearing of the speaker and on whether the addressees are accustomed to receiving commands from that particular speaker. However, it seems clear that in the case of legal norms – for example, laws issued by a legislative authority – there is no such personal relationship. On the contrary, there is a sharp separation, not only physical but temporal, between those who could be conceived as parties to that relationship. Indeed, the relevant relationship here could be, in fact, between the imperative and the addressee, not between the sender and the addressee. For all intents and purposes, here, the only role of the sender has been to attach a certain label ("law") to a certain utterance following a certain process, at which point that utterance functions with complete independence or consideration in relation to the sender. Hence, they are "independent imperatives": they function as imperatives regardless of who the active subject is, and this is because they have been declared as "law", a label that directs the addressees to establish a connection between the stated pattern of behaviour and a feeling of constraint towards the performance of the action.

[111] As we shall see in point IV.2.4 below, this function of the artefact can be achieved in the domain of the interpretation of legal texts, as long as the interpreters of the legal text act with the presupposition of intention.

[112] See e.g. Olivecrona, *Law as fact,* 2 ed.

[113] For further development and discussion, see e.g. Torben Spaak, *A Critical Appraisal of Karl Olivecrona's Legal Philosophy* (Cham: Springer International Publishing, 2014); Julieta A. Rabanos, “La máquina del derecho y sus engranajes. Karl Olivecrona sobre derecho, autoridad, y normas jurídicas como imperativos independientes”, *Analisi e diritto* 21 (2021): 145-177.

With this approach, on the one hand, Olivecrona removes the importance of the sphere of emission (the legal authority) and places all attention on the sphere of reception (the listener or reader), explicitly recognising that it is the addressees who, in any case, define the content and scope of that which is labelled "law" (without necessarily involving conjectures about the intention of the issuer). On the other hand, it not only dispenses with fiction about the legislative intention but also provides a formal element of explanation about the imposition of the label "law" on a given text. All this while maintaining the explanation of in what sense the text labelled "law" is (the formulation of) an imperative, in what sense it functions as an imperative, and in what sense it is to be understood as a normative enunciation.

One possible concern regarding this approach, among others[114], might be that it could invite confusion and disagreement regarding the meaning or content of "laws" – that is, that such an approach leaves the door open to the possibility of "laws" ultimately not having some kind of determinate or determinable meaning. However, this worry does not apply here for various reasons. Firstly, the concern might arise from the idea that Olivecrona does not (sufficiently) concern himself with the study of legal interpretation and its consequences. There are several possibilities here. First, Olivecrona might have an implicit cognitivist approach to legal interpretation. Thus, for him, words have intrinsic, essential and objective meanings to be, in some sense, *discovered* by the readers. However, Olivecrona argues for the opposite view: the reader (especially if deciding an individual case) must continually evaluate the legal text to decide what it means, and evaluations are not objective.[115] Secondly, Olivecrona might have considered the answer to this topic too commonplace or too commonsense for him to explicitly tackle, where the answer is that words have a determinate or determinable meaning depending on the relevant community of readers. This is plausible, as it is coherent with his explicit views about the kind of evaluations when classifying a text as "law". On the one hand, the readers might contingently agree on the evaluations because of their shared background, but there is never certainty about these agreements and whether they will hold at specific times and in specific cases. On the other hand, these evaluations are never completely fixed, always experiencing modifications.[116] Finally, Olivecrona might have considered this topic to be outside what he endeavoured to explain, which is the relation between imperatives and addressees and the mechanism through which the imperative directs the addressee's behaviour – in this context, words have the meaning that the individual addressees give them, and depending on that, they direct or not their behaviours. This is also plausible, and also perfectly compatible with the second possibility we have just analysed.

[114] See point IV.2.4 *in fine*.

[115] Olivecrona, *Law as fact,* 2 ed, pp. 212-215. For further discussion and critical analysis, see e.g. Spaak, *A Critical Appraisal of Karl Olivecrona's Legal Philosophy*, pp. 182 ff.

[116] Olivecrona, *Law as fact,* 2 ed, p. 215.

On the other hand, this concern might spawn from the belief that approaches such as Olivecrona's would need to provide an account about how confusion and disagreement regarding the meaning or content of "laws" could be avoided or fixed, or otherwise considered to be as non-existent or a non-issue. In other words, this belief seems to presuppose that Olivecrona's approach would need to be concerned with how things *ought* to be – how to conceive legal phenomena in a way in which legal texts' addresses are not confused or disagree about their meaning or content – and not (only) with how things *are* – how legal texts have meanings without reference to any intention related to the creation of those texts. However, is this belief justified, and is this an issue for Olivecrona's approach? It seems not. On the one hand, it is quite difficult to see the reasons why a theoretical approach that endeavours to describe and explain legal phenomena as they are would also need to concern itself with how legal phenomena ought to be (or how they ought to be described and explained to account for a particular hypothesised desirable image of them). On the other hand, the fact that addresses' confusion and disagreement about the meaning or content of "laws" might exist is something that Olivecrona's approach indeed accounts for and does not suppose an issue for it.

IV.2.4. The Intentional Reader

One example of the third strategy, claiming that it is impossible to find anything analogous to the "speaker's intention" of a legal author in a legal context as we could do it with an ordinary author in ordinary conversation, so the attention must be put not on the speaker but on the reader, is Francesca Poggi's approach.[117]

While analysing the plausibility of applying Grice's conversational model to the analysis of legislative interpretation, Poggi denies the possibility mainly because there is no "legislative intention" regarding the meaning of the output to be found – at least, not in the relevant Gricean sense of a "speaker's intention". She thoroughly analyses different attempts to identify or reconstruct the "legislative intention" and rejects all of them. In a very brief fashion, she rejects the summative accounts because of their difficulty in identifying the subjects whose intentions would be aggregated and whose solution – identifying them with the majority that approves a certain text – clashes with the legal fact that decisions are imputed to the entire legislature and not only to the majority. Next, she rejects the collective intentionality thesis of Searle and Bratman-Ekins. The first one, although it solves the second problem of summative accounts, has the problem that it does not allow the identification of a collective intention to approve a text X expressing a norm N. The second one has the same problem: it fails to show that the legislative intention (as they reconstruct it) includes the intention to imply something more than the approved text, i.e. a certain norm N. Finally, she rejects counterfactual models because of the difficulty of reconstructing a univocal counterfactual will, because of the controversial status of the logic of counterfactuals, and because counterfactual intention could not play the same role as "speaker's intention" since it is not "real".[118]

[117] See Poggi, *Il modello conversazionale*. For further development and discussion, see e.g. Lorena Ramírez-Ludeña, "La interpretación jurídica y el modelo conversacional. Algunas reflexiones sobre *Il modello conversazionale* de Francesca Poggi", *Analisi e diritto* 23 (2023): 11-24; Julieta A. Rabanos, "Dos comentarios a *Il modello conversazionale*, de Francesca Poggi", *Analisi e diritto* 23 (2023): 41-58.

In this sense, Poggi finds that an "intention" of producing a text can be identified, but not an intention of producing a certain norm (a meaning of that produced text). She says: "legislation is an intentional phenomenon (...) because it is the (intentional) product of intentional human acts", although this intention "is not a communicative intention in the Gricean sense, or rather, it is not the intention to communicate something and not something else, but, instead, it is the shared intention to participate in a procedure by accepting its rules and accepting that what results (...) will be emanated as, and will have the value of, law."[119] Hence, the maximum that could be reached through the reconstruction of the "legislative intention" would be the collective intention to approve a certain text X; not to approve a certain text X that expresses a certain norm N. This intention would be sufficient, according to Poggi, "to establish the assumption of normativity and legality of the texts of law": that is, that the approved text expresses norms, and in particular legal norms.[120]

However close this claim can be to Olivecrona's, Poggi explicitly rejects the possibility of adopting a position that suppresses intention to such an extent. In her words, our intuitions about norms being the expression of the will or intention of someone are too strong to do entirely away with them in legal contexts. However, Poggi's rejection comes from the context of utterance and the intuitions of the practice participants (mainly, it seems, the possible addressees and/or interpreters). Here, there can be a shared assumption that any utterance coming from a specific body and taking certain forms must be understood as normative[121] – one assumption that seems almost absolute among the interpreters.[122]

Would this last rejection imply that Poggi's approach contradicts itself, as it ultimately would accept some reference to the legislator's intentions as necessary? The answer is negative. As we have seen, Poggi argues that it is possible to find something akin to "speaker's intention" in law regarding the intention of approving a certain text X, but it is impossible regarding the intention of approving a certain meaning A of text X (i.e. a certain norm N). She then argues that: (1) even if no real intention of creating norms can be found in the acts of legislative authorities, we – the participants within legal contexts' practices – are so convinced that this norm-creation intention exists or that we need that intention to undertake legal interpretation that we just *assume* it exists; (2) this practice revolves based on that assumption, which constitutes a cornerstone of the practice and a background assumption of every act within it;

[118] Poggi, *Il modello conversazionale*, pp. 230 ff. A brief overview of her arguments on this particular point can also be found in Francesca Poggi, "Against the conversational model of legal interpretation", *Revus* 40 (2020): 9-26.

[119] Poggi, *Il modello conversazionale*, pp. 320-321.

[120] Poggi, *Il modello conversazionale*, pp. 321, 324.

[121] Poggi, *Il modello conversazionale*, p. 235.

[122] See Pierluigi Chiassoni, "Significato letterale: giuristi e linguisti a confronto", in Vito Velluzzi (ed), *Significato letterale e interpretazione del diritto* (Turin: Giappichelli, 2000): 1-63, p. 52.

and, thus, (3) we cannot describe or explain the practice (as it is) without referring to this contextual assumption of existence of norm-creation intention - even if that intention does not exist, even if it is not possible to find it. This marks a clear distinction with Olivecrona's position, which seems to assume that we can describe or explain legal phenomena or practices without any reference to any legislative intention, even if that reference is to ideas that addressees have about it – and that, even more, we do better that way because we do away with "dangerous fictions". On this point, Poggi's approach might be closer to Fish's ideas than Olivecrona's.

IV.2.5. Dispensing of intention

At this point, we want to inquire whether the three strategies and/or their constructive combination allow us to circumvent the issues caused by the possibility of texts being generated by agents lacking intentions. The common denominator of these strategies is that they downplay the intention of the author(ity) in favour of the intention of the reader or the community of readers. If we add the idea that legality is conferred to utterances in the impersonal process of producing law, we have all the necessary elements for addressing the problems that Alexander thought the elimination of intention entails.

On the side of the product brought into being by those agents, the legal character of the products does not depend on the intention of the author of the artefact. We have argued that there are good reasons to detach an object's artefactual character, which can be understood as the product of its users' intentions. For a textual, oral, or other output to be law, it is enough that it performs its proper function assigned to it by its subjects. In this sense, it is entirely optional whether the subjects or users will have an idea of a real or imagined author behind the utterances as long as the function of the utterances is consistently maintained.

The authority, strength and content of the artefact that performs as law is made possible because its creation fits the requirements placed on the production of law within a normative system. One of those requirements might be that we should count only the content of those utterances made by an intentional agent as law, but it, of course, needn't be. While it will often be the case that an authority attaches the label law to its textual product, the very act of attaching this label is not constitutive of the legal character of utterances. In other words, the potential of artificial intelligence to act as a legal authority is not necessarily undermined by the lack of intention on its part. However, the side effect of this move might be that the content of the utterances might seem indeterminate to an unbearable degree.

However, the form of the artefact that can, in fact, be created by an author lacking intention, in the position of an authority lacking intention, might be determined regardless of the determinacy of its content. The content of the utterance that we consider legal is determined by acts of ascribing normative content to the utterances and not by the intention of the intentional author, backed by an intentional authority. After all, this is what judges do when they interpret legal texts produced by supposedly intentional authors. They choose among the possible meanings of a text in order to ascribe the normative meaning according to their (often reasoned) ideas about the deontic positions of the subjects. An intention, or intention in general, real or imagined, can play a role in this determination. Still, it's certainly not necessary for the text to give rise to a determinate normative legal meaning.[123]

In this way, while we might still imagine an intentional author behind utterances, this imagination is certainly optional for the artefactual character of law, its authority or the determination of its content. The recipients of any text are namely "fully equipped" to 1) determine the use functions of the text, 2) confer the necessary legal authority to its maker or generator, and 3) determine the normative content of the text. We are still free to indulge in the intuition, assumption, and idea that norms are expressions of the will of intentional agents as long as we are aware that this intuition is not necessary to explain the artefactual character of law, the authoritative character of legal texts, nor the normative content of those texts.

## V. SOME CONCLUDING REMARKS

In conclusion, all the dilemmas about agency, intelligence, and the inner workings of generative AI did prompt our previous inquiry. Still, they might have needlessly complicated the problem, especially for philosophers of law who are often not in tune with contemporary technological developments. In this final part, we propose a thought experiment that mimics the situation in technology that the law is confronted with and that will most certainly persist in future philosophical discussions about law AI-generated law. It goes as follows:

> A tribe without any formal regulation of their social life finds a book containing only prescriptive sentences. For some reason, the members of the tribe can perfectly well understand the writing in the book. They read the writing, have a tribal meeting, and decide that the writing will, from now on, be used to regulate their social life.
>
> The tribe goes on to conduct its life based on the writing in the book. The chieftain decides disputes based on the text of the book; the kids read the book to learn how to behave. The book is consulted in all cases in which there is a doubt about which course of action to take, and behaviours are judged based on the adherence or defiance of the rules expressed by the text.
>
> Shortly thereafter, one of the tribe members discovers deep in the woods a field with a thousand cats (including Marija and Irina) merrily stepping and jumping on a thousand typewriters that leave letter marks on paper. A person is overlooking their "work" (taking exceptional care of the imagined cats, feeding and petting them if and when they feel like it). Whenever the cats manage to type something that sounds sensible to the person, the person puts it down on a separate piece of paper. When the amount of sensible writing reaches 150.000 characters, he binds the paper in a book and leaves it among the threes.
>
> Dumbfounded by the discovery, the tribe member who witnessed the extraordinary sight returns to the village and announces it to the others by ringing a bell and shouting: "*Law is dead. We have no law.*"[124]

[123] See Bojan Spaić, "Institutional Control of Interpretation", in Mortimer Sellers, Stephan Kirste (eds), *Encyclopedia of the Philosophy of Law and Social Philosophy* (Springer, 2024), pp. 1-6.
[124] Unintentionally paraphrasing the madman from *The Gay Science* (Friedrich Nietzsche, 2018).

The tribal rulebook containing prescriptive sentences is not created with the intention of making law. With all the research on the cognitive capacities of nonhuman animals[125], and with cats having a more (Marija) or less (Irina) pronounced tendency to be agents and to influence the behaviour of their owners, there is no authorial type intention in the cats to produce the legal text. The authorial intention cannot be borrowed from the human editor because 1) she is not an author of the writing, and 2) her intention is not to edit a legal text. Consequently, there is no intention of cats to create any legal rulebook that would regulate the tribal life, to produce a codification, nor is there an intention on the part of the cats or the humans to influence people by way of rules to behave in a certain way.

Our thought experiment seems to point towards the idea that the legal character of the outputs of artificial intelligence is not premised on the existence of intention behind a text that supposedly expresses legal norms. Either because it is an artefact that needs no intention to be created (strategy one), because its existence and functions do not depend on the intention of any creator (strategy two), and/or because it is enough for the readers to collectively share the assumption that some outputs that come to be from a specific source and from a particular process necessarily have certain characteristics (strategy three). If the source looks like law, is intelligible as law, and is recognised as law, then it probably is law – right?

For all these reasons, then, our central question finally has its answer: outputs of LLMs *can* count as law in all three of our scenarios on the condition that their function is legal. The death of the legal author – at least as we traditionally knew it – seems inexorable: it has run down the curtain and joined the choir invisible.[126] And now, the normative question of whether AI outputs *should* be recognised as law – whether we should leave the legal author cease to be – can begin to be adequately answered.

[125] See Herbert L. Roitblat, "Animal Cognition", in William Bechtel & George Graham (eds), *A Companion to Cognitive Science* (London: Wiley, 2017), pp. 114-120.

[126] Monty Python, *Dead Parrot* (1969): https://www.youtube.com/watch?v=4vuW6tQ0218

**ACKNOWLEDGEMENTS**

This paper results from research conducted within the Horizon Twinning project "*Advancing cooperation on The Foundations of Law – ALF*" (project no. 101079177). The project is financed by the European Union.

Initial versions of the paper were presented within the following conferences: *Belgrade Philosophy of Law Week conference on Artificial intelligence and Philosophy of Law*, organized by the Center for Legal Fundamentals of the Faculty of Law University of Belgrade on 18th of November 2023; *Jurix2023 Workshop on AI, Law and Philosophy* on the 18th of December 2023 at the University of Maastricht organized by Jaap Hage, Ludi van Leeuwen, Bart Verheij, and Antonia Waltermann. We are grateful to the organizers and the participants for valuable comments on the initial versions of the paper.